\documentclass{webofc}

\usepackage[varg]{txfonts}   
\usepackage{hyperref}
\usepackage{url}
\hypersetup{colorlinks=true,citecolor=blue,urlcolor=blue,linkcolor=blue}
\begin{document}
\title{Hadronic Molecules: Theory }
%
%

\author{\firstname{Christoph} \lastname{Hanhart}\inst{1}\fnsep\thanks{\email{c.hanhart@fz-juelich.de}} }

\institute{Institute for Advanced Simulation
and Cluster of Excellence ``Color meets Flavor'', Forschungszentrum J\"ulich, D-52425 J\"ulich, Germany
          }

\abstract{Since the turn of the century it becomes evident that the naive quark model is not capable to describe the rich phenomenology of hadrons, especially in the doubly heavy sector. For a large number of those states a molecular structure appears to be a natural description. In these proceedings I review the theory of hadronic molecules, how they can be described systematically in effective field theories and what imprint this structure leaves in observables. On two examples I explain non-trivial predictions emerging from the molecular hypothesis that can be tested in future experiments and lattice QCD calculations, promising deep insights into the inner workings of QCD.
}
\maketitle
\section{Introduction}

This section is based on Refs.~\cite{Guo:2017jvc,Hanhart:2025bun}.
Hadronic molecules are multi-quark states, whose substructure is given by color neutral hadrons.
Thus, atomic nuclei made of nucleons, protons and neutrons, fall into this category.
There is no striking evidence that, e.g., a six quark component in needed in deuteron wave function.
 Binding energies $E_B$
 (for larger nuclei normalised by
 the number of consitutents)
 for nuclei range from 0.13~MeV for the hypertriton, to about 10~MeV for heavier nuclei.  
We expect similar values for other hadronic molecules.

To provide some theoretical understanding, we may study the scattering of two-particles feeling a zero range interaction---this is justified
as soon as the de Broglie wave length connected to the bound system, $\lambda_b=1/\gamma$, where the binding momentum
is defined by 
\begin{equation}
\gamma = \sqrt{2\mu E_B} \quad \mbox{with} \quad E_B=m_1+m_2-m_B ,
\end{equation}
 is significantly larger than the 
range of forces, in what follows denoted by $R$.  Then the scattering amplitude at tree level in momentum space is given by a constant 
and we get for the loop expansion of the scattering amplitude in non-relativistic kinematics~\cite{Guo:2017jvc}
\begin{equation}
T=C \sum_{n=0}^{\infty}(C \Pi(E))^n =  \frac{C}{1+C \Pi(E)} 
\label{eq:TNR}
\end{equation}
where 
\begin{equation}
\Pi(E)= -\int \frac{d^3 q}{(2\pi)^3}\frac{F^2_\Lambda(q)}{E-q^2/(2\mu)+i\epsilon} = \frac{\mu}{2\pi}
\left(\Lambda+i\sqrt{2\mu E+i\epsilon}\right) + {\mathcal O}(\Lambda^{-1}) \ ,
\end{equation}
where $F_\Lambda(q)$ is some regularisation function and $\Lambda$ the regularisation scale.
Here 
 $\mu{=}m_1m_2/(m_1{+}m_2)$ denotes the reduced
mass of the external particles. The condition for the appearance of a near threshold pole
in the amplitude may thus be expressed as
\begin{equation}
C^{-1} = -\Pi(-E_B)  = \frac{\mu}{2\pi}
\left(-\Lambda + \gamma\right) + {\mathcal O}(\Lambda^{-1}) \ .
\end{equation}
With this
the renormalised scattering amplitude takes the simple form
\begin{equation}
T=\left( \frac{2\pi}{\mu}\right)\frac{1}{\gamma+i\sqrt{2\mu E+i\epsilon}} \ ,
\label{eq:mol_boundren}
\end{equation}
which depends, besides on the energy, only on the binding energy and the masses of the external
particles. In particular one gets for the residue at the pole in the non-relativistic normalisation used here
\begin{equation}
g^2=\lim_{E\to -E_B}(E+E_B)T_{\rm NR}=\frac{1}{d\Pi(E)/dE|_{E=-E_B}}=\frac{2\pi\gamma}{\mu^2} \ ,
\label{eq:molres}
\end{equation}
also independent of the detailed dynamics that lead to the appearance of the pole. Because of this shallow bound states develop
universal properties---for a review see Ref.~\cite{Braaten:2004rn}.

\begin{figure}[t]
	\centering
	\includegraphics[width=0.8\textwidth]{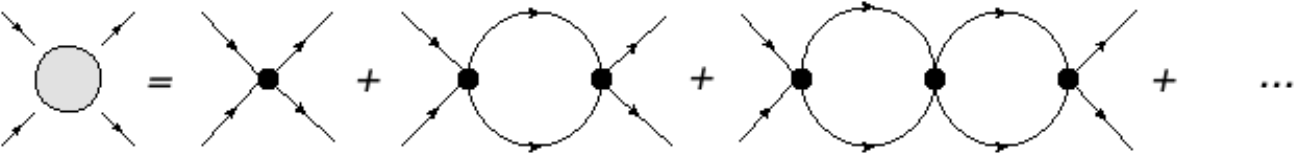}
	\caption{Most simple scattering equation, where the interaction is given by a $\delta$-distribution in $r$-space/a constant in momentum space.
	\label{Fig:LSe}}
\end{figure}

The discussion above applies to purely dynamically generated states. The equations were generalised by Weinberg~\cite{Weinberg:1965zz} (inelastic channels were later included in Ref.~\cite{Baru:2003qq}). 
The study starts from a two component wave function of a bound state and the Hamiltonian
$\hat {\mathcal H}$,
\begin{equation}
|\Psi \rangle = \left(\lambda|\psi_0\rangle\atop 
\chi (\mathbf{p}) |h_1h_2 \rangle \right) 
\qquad \mbox{solving} \qquad \hat {\mathcal H} |\Psi\rangle = E |\Psi\rangle,~~
\hat {\mathcal H} = \left(\begin{array}{cc}
\hat{H}_c&\hat{V}\\
\hat{V}&\hat{H}_{hh}^0
\end{array}
\right) \ .
\end{equation}
Here $\psi_0$ ($h_1h_2$) is the compact (two hadron) component of the wave function. 
In the Hamiltonian the hadron-hadron interaction was removed by some field redefinition~\cite{Weinberg:1962hj}\footnote{This is possible as long
as there is not additional pole very close to the threshold~\cite{Baru:2010ww,Hanhart:2011jz,Guo:2016bjq}.}. In this way the information
on the nature of the state gest transferred to the coupling of the physical state to the continuum as detailed below.
The quantity of interest here is 
\begin{equation}
\lambda^2 = \langle \psi_0|\Psi\rangle ^2  \ ,
\end{equation}
which is the probability to find the compact component of the wave function in the full wave function
and is thus a direct measure of the composition of the wave function.
The normalisation condition for physical bound states then allows one to relate
$\lambda^2$ to the transition strength of the physical state to the continuum at 
zero momentum transfer, which is in fact the unrenormalised channel coupling of the
state:
\begin{equation}
g_0^2 = \frac{2\pi\gamma}{\mu^2}\left(\frac{1}{\lambda^2}-1+\mathcal{O}\left(R\gamma\right)
\right) \ .
\label{eq:g0oflambda}
\end{equation}
It is straightforward to calculate the scattering amplitude within the same formalism under the assumption
that the scattering is dominated by a single pole. All it takes is to replace in Eq.~(\ref{eq:TNR}) $C$ by
$g_0^2/(E-M_0)$. Now the scale dependence is absorbed into the bare mass $M_0$ via the
renormalisation condition $E_B=-M_0+g_0^2\mu/(2\pi)(\Lambda-\gamma)$, where we used that 
the analytic continuation of the square root of the energy on the first sheet (here the focus is on bound
states) gives the positive imaginary momentum.
Then we find for the scattering matrix~\cite{Guo:2017jvc}
\begin{equation}
T=\frac{g_0^2}{E+E_B+g_0^2\mu/(2\pi)(ip+\gamma)} \ .
\label{eq:TNRpole}
\end{equation}
For a pure molecule we have $\lambda\to 0$ and accordingly $g_0^2\to\infty$. In that limit  
Eq.~(\ref{eq:TNRpole}) agrees with Eq.~(\ref{eq:mol_boundren}).
Moreover, the residue at
the pole is $g_0^2$ multiplied with the $Z$-factor that happens to agree with $\lambda^2$. This
gives
\begin{equation}
g(\lambda^2)^2 = \lambda^2 g_0^2 = \frac{2\pi\gamma}{\mu^2}\left(1-{\lambda^2}+\mathcal{O}\left(R\gamma\right)\right) \ ,
\end{equation}
which agrees to Eq.~(\ref{eq:molres}) for $\lambda^2=0$, when the range corrections are omitted. Eqs.~(\ref{eq:g0oflambda})
and~(\ref{eq:TNRpole}) nicely illustrate what the Weinberg criterion really does: It measures the importance of the
term non-analytic in the energy, $p=\sqrt{2\mu E}$, which appears, since the two-hadron intermediate state can go
on shell and is thus specific for the molecular component, relative to the terms analytic in energy that also appear
for compact structures.
There are a large number of works available in the literature that discuss generalizations of Weinberg's
criterion also to virtual states and resonances~\cite{Baru:2003qq,Matuschek:2020gqe,Bruns:2019xgo,Oller:2017alp,Kang:2016jxw,Sekihara:2016xnq,Kamiya:2016oao,Guo:2015daa,Sekihara:2014kya,Aceti:2012dd,Gamermann:2009uq,Kinugawa:2024kwb}, which, however, we cannot discuss in detail here.

It can be shown that potential scattering leads to a positive effective range~\cite{Bethe:1949yr}---in other words: a positive
effective range is an unambiguous signature of a purely molecular state~\cite{Weinberg:1965zz,Esposito:2021vhu,Baru:2021ldu,Li:2021cue,Albaladejo:2022sux}.
 However, not even a sizeable   negative effective range
provides a unique signature of a compact structure, since coupled channel effects also induce 
a negative contribution to the effective range~\cite{Baru:2021ldu,Matuschek:2020gqe,Esposito:2021vhu}.

As pointed out above the residue at the pole of a physical state gets maximal, if the state
is a pure molecule. For shallow bound states this statement can be turned into a prediction
for the residue. However, there are corrections that scale as $(R\gamma)$. 
Thus, if the binding of a state gets too deep ($\gamma$ gets large), 
the corrections can be very large. However, we anyhow assume from now on that 
because of the large coupling loops appear at leading order for hadronic molecular states,
even if we cannot give a quantitative prediction for the coupling, whenever the state of interest is
either a resonance or too deeply bound.

In the rest of these proceedings two examples are discussed briefly that illustrate
how symmetries and their breaking help us to go beyond the general concepts
introduced above:
 \begin{itemize}
\item Example I: Implications of Heavy Quark Spin Symmetry (HQSS), illustrated on an
exploratory study of the $1^{--}$ system and some spin partners, based on Ref.~\cite{Dong:2026fdi}.
\item Example II: Isospin Symmetry and its breaking, illustrated on
the example of a chiral EFT for $X(3872)$ and isovector partners, based on Ref.~\cite{Ji:2025hjw}.
 \end{itemize}

\begin{figure}[t]
	\centering
	\includegraphics[width=.7\textwidth]{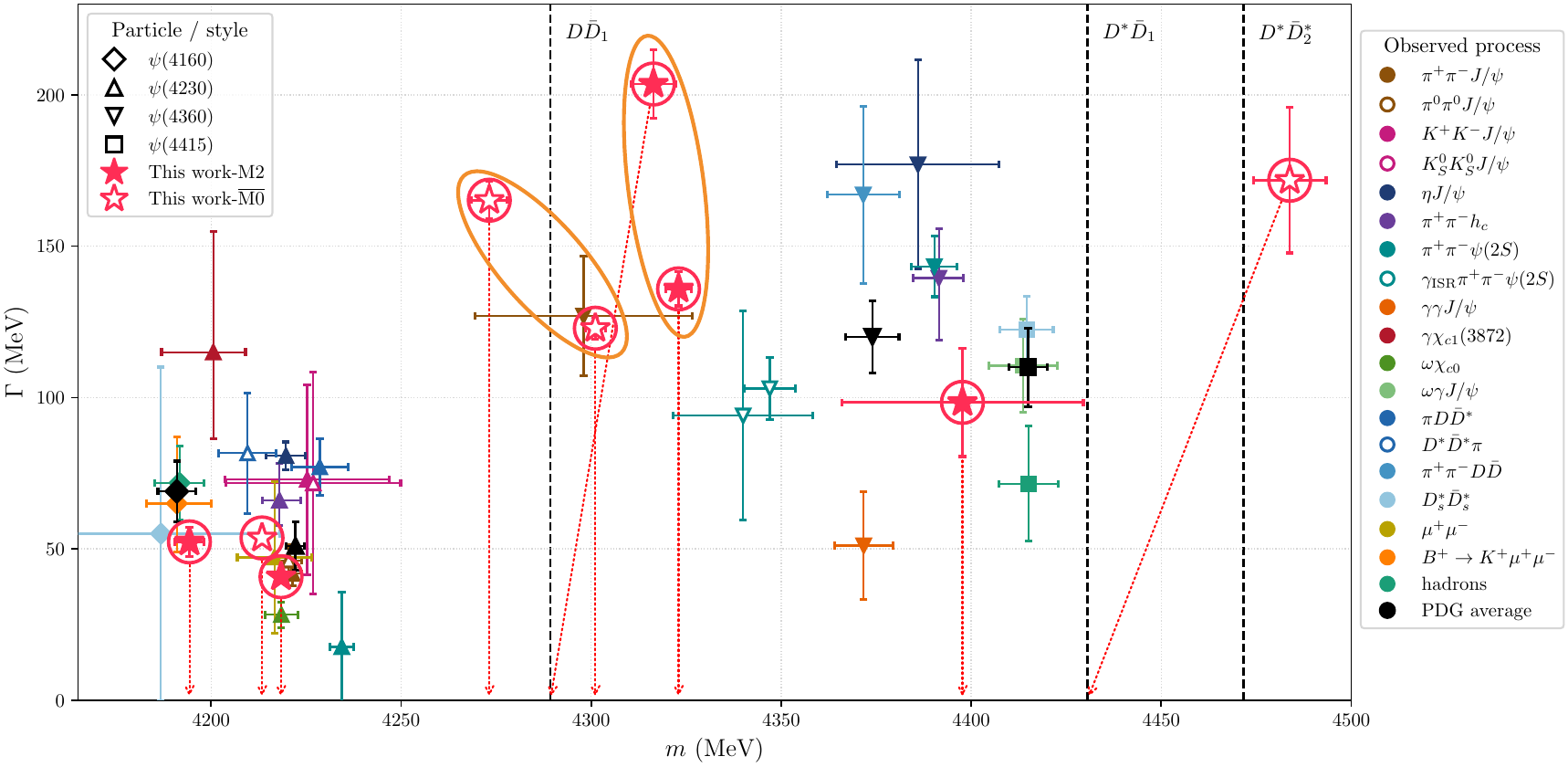}
	\caption{Comparison of pole locations deduced by the different experiments to those determined in 
	two different fits of our analysis. Open (filled) starts refer to the poles extracted from Fit $\overline{\mbox{M0}}$ (Fit M2).
	 The two poles encircled by the ellipses refer to the same state.  
	Details in the main text. The data are from Refs.~\cite{BESIII:2023wsc,LHCb:2013ywr,BES:2007zwq,BESIII:2022kcv,BESIII:2023cmv,BESIII:2022qal,BESIII:2022joj,BESIII:2021njb,Ablikim:2020jrn,BESIII:2020oph,BESIII:2019gjc,BESIII:2018iea,BESIII:2019qvy,BESIII:2023tll,BESIII:2022quc,BESIII:2021yal,BESIII:2016adj,Belle:2014wyt,BaBar:2012hpr,BESIII:2026qhe,BESIII:2024jzg}.
	\label{Fig:Yspectrum}}
\end{figure}

\section{Example 1: Exploratory study of the $1^{--}$ system }

HQSS is an exact symmetry of QCD in the limit of heavy quark masses going to infinity. It emerges,
since in this limit the spins of the heavy quarks decouple and accordingly both the heavy quark spin and the light quark 
total angular momentum get conserved individually. In Ref.~\cite{Dong:2026fdi} we demonstrated that exclusive
decay data for $e^+e^-~\to$~hadrons  in the mass range from 4.1 to 4.6~GeV  can be understood from an interaction
constrained by HQSS.

To be concrete, we studied the scattering of the spin doublets $\{D,D^*\}$ with $j_\ell=1/2$ and $\{\bar D_1,\bar D_2^*\}$ with $j_\ell=3/2$
(combined into isoscalars with $J^{PC}=1^{--}$). Thus, $j_\ell^{\rm tot}\in \{1,2\}$ and we get in total four possible short ranged operators~\cite{Guo:2017jvc},
which read
\begin{eqnarray}\nonumber
  \langle{s_Q,j_\ell^{\rm [tot]},J}|\mathcal H_{\rm int}|{s_Q',j_\ell^{{\rm [tot]}'},J}\rangle {=} 
\langle{\overline{s_Q,j_\ell^{\rm [tot]},J}}|\mathcal H_{\rm int}|{\overline{s_Q',j_\ell^{{\rm [tot]}'},J}}\rangle
{=}\delta_{s_Qs_Q'}\delta_{j_\ell^{\rm [tot]} j_\ell^{{\rm [tot]}'}}F_{j_\ell^{\rm [tot]}}^d & & \hspace{-0.5cm} , 
\\
\langle{s_Q,j_\ell^{\rm [tot]},J}|\mathcal H_{\rm int}|{\overline{s_Q',j_\ell^{{\rm [tot]}'},J}}\rangle {=}
\langle{\overline{s_Q,j_\ell^{\rm [tot]},J}}|\mathcal H_{\rm int}|{s_Q',j_\ell^{{\rm [tot]}'},J}\rangle{=}
\delta_{s_Qs_Q'}\delta_{j_\ell^{\rm [tot]} j_\ell^{{\rm [tot]}'}}F_{j_\ell^{\rm [tot]}}^c & &\hspace{-0.5cm} , 
\end{eqnarray}
where we changed the basis to $|D^{(*)}\bar D_{1,2}^{(*)}\rangle {\Leftrightarrow} |{s_Q,j_\ell^{\rm [tot]},J}\rangle$ and 
$|D_{1,2}^{(*)}\bar D^{(*)}\rangle {\Leftrightarrow} |\overline{s_Q,j_\ell^{\rm [tot]},J}\rangle$. Out of those only three are operative
in the vector channel. Thus the interaction in the three-channel system $D\bar D_1$, $D^*D_1$ and $D^*D_2^*$ is controlled by
3 parameters which we chose complex to account for the large number of possible inelastic channels. 

In the simplest realisation of the model, Fit M0, we only include the scattering channels mentioned above to fit the data 
of the final states $J/\psi \pi\pi$ (including the differential $J/\psi \pi$ spectra), $h_c\pi\pi$, $D\bar D^*\pi$, $D^*\bar D^*\pi$, $J/\psi\eta$ and $\chi_{c0}\omega$.
 In addition to the mentioned parameters we need parameters to control the production, the transitions to the non-elastic final states and the $\pi\pi$
 final state interactions---in total Fit M0 has 37 parameters to fit 1240 data points. It turns out that fitting the data as they are, the fit fails to describe 
 structures in the $h_c\pi\pi$ and $D\bar D^*\pi$ data around 4.45~GeV. However, increasing the weight in the fit of these data sets by hand in the
 fits, generating Fit $\overline{\mbox{M0}}$, allows for a good overall description of the total cross section. The poles from the latter
 fit are shown by the open star symbols in Fig.~\ref{Fig:Yspectrum}. The overall best fit is achieved, if we include in addition two compact pole terms,
 identified with $\psi(4160)$ and $\psi(4415)$ in Fit M2,
 which is controlled by 61 parameters.
 The resulting poles are shown as filled stars in Fig.~\ref{Fig:Yspectrum}. Once we extend the study to include also the $P$-wave $D^{*}\bar D^{*}$ channels,
  we expect a more conclusive view on the role of the compact $\bar cc$ states. These channels were already considered in Ref.~\cite{Nakamura:2023obk},
 where, however, no spin symmetry constraints were used.
 
 In both fits mentioned above we observe large channel couplings driven by the spin-symmetric scattering potential. The most striking
 effect of these is that the state around 4.3~GeV appears as a pair of poles on different Rieman sheets that influence the data in strongly
 in different energy ranges. This type of pole pair is not to be confused with the famous two-pole structures (for a review see Ref.~\cite{Meissner:2020khl}):
 In the latter case a single structure in experiments is interpreted as emerging from two distinct states represented by poles on the same Rieman
 sheet. In contrast to this we have here a pair of poles on different Rieman sheets referring to the same state. 
 
 The three parameters fixed in the vector channel allow us to also make predictions for some of the possible spin partners
 (for a complete study, based on modelling the interactions, see Ref.~\cite{Ji:2022blw}): We find evidence for bound
 $0^{-+}$ and $3^{--}$ states, while the exotic $0^{--}$ state turns out badly determined, with some preference for a pole
 on the unphysical sheet (virtual state).
 
 To summarise this section we stress that the $1^{--}$ channel is very rich and it is yet to be resolved what the true spectrum
 of vector states is. In any case it would be desirable to also have data for quantum numbers other that the vector channel,
 since the existence and the location of the spin multiplets will provide additional non-trivial insights especially on the role
 played by HQSS and its breaking.

\section{Example 2: Chiral EFT for $X(3872)$ and its isovector partners}

Isovector partners of $X(3872)$ were already predicted to exist almost degenerate with the $X$ in tetraquark studies shortly after the discovery of 
the state~\cite{Maiani:2004vq}---see Ref.~\cite{Maiani:2020zhr} for a study on $B$ decays. This prediction survived various refinements of the model. 
Until recently a prediction for such a partner
state was not possible model independently within the molecular picture, since the hadron-hadron interaction in the isospin 1 and the isospin 0 channels are 
a priori independent. The situation changed when LHCb published very high precision data for $B^+\to K^+ J/\psi\pi^+\pi^-$~\cite{LHCb:2020fvo}, including 
the corresponding $\pi^+\pi^-$ invariant mass distributions~\cite{LHCb:2022jez}. Especially the latter data allow one to disentangle the $\omega J/\psi$
and $\rho J/\psi$ distributions which provides insights into the isospin composition of the $X$ signal.

As detailed above, in leading order a molecular state is described by the elastic scattering of its constituents with a constant 
interaction strength~\footnote{In a high precision study this is to be supplemented by a one-pion exchange following Refs.~\cite{Baru:2011rs,Du:2021zzh},
since the three-body $D\bar D\pi$ cut cannot be absorbed into any effective parameter. }. To describe the data there must be transitions
from the $X$ structure to both, isospin 1, $\rho J/\psi$, and isospin 0, $\omega J/\psi$, final states~\cite{LHCb:2022jez}. This picture 
can be refined further by a direct measurement of the three pion rate (and line shape) from $X\to J/\psi\pi^+\pi^-\pi^0$ predicted in Ref.~\cite{Dias:2024zfh}. 
One large isospin violating contribution is the mass difference
of the neutral and charged $D\bar D^*$ states. It turn out, that this alone---in combination with the large differences in effective phase spaces
for the two $VJ/\psi$ channels that originate from the vastly different widths of the two vector mesons---that was claimed 
to capture the $X$ phenomenology successfully~\cite{Gamermann:2009fv},  is not sufficient to describe the new high quality data.
In fact, if we impose isospin symmetry of the couplings consistent with the established hadron phenomenology,
it turns out that a significant isospin mixture in the molecular picture needs a strongly attractive interaction also in the
isospin 1 channel~\cite{Dong:2026qbr}.
Due to this additional interaction the scattering amplitude
has in addition to the pole of the $X(3872)$ located at~\cite{Ji:2025hjw}.
\begin{equation}
 {E_X=\left(-160^{+57}_{-74}-125^{+23}_{-38}\,i\right) \rm keV}
     \label{eq:Ex}    
\end{equation}
below the $D^0\bar D^{*0}$ threshold
on the physical sheet of that channel, 
an additional pole at
\begin{equation}
E_{W_{c1}}=\left(3.1\pm 0.7+ 1.3^{+1.9}_{-0.6}\,i\right)\ \rm MeV
\end{equation}
above the $D^+D^{*-}$ threshold---this pole was predicted to exist in Ref.~\cite{Zhang:2024fxy}; further evidence comes from 
lattice QCD~\cite{Sadl:2024dbd}. 
It is located on the $+-$ Rieman sheet, which connects to the physical axis only
at the $D^+D^{*-}$ threshold.
Accordingly this pole can only show up as a cusp.
Note that this location of the pole does not come by chance, but can be understood
quite straightforwardly~\cite{Dong:2026qbr}.

\begin{figure}[t]
\centering
\sidecaption
\includegraphics[width=9.5cm,clip]{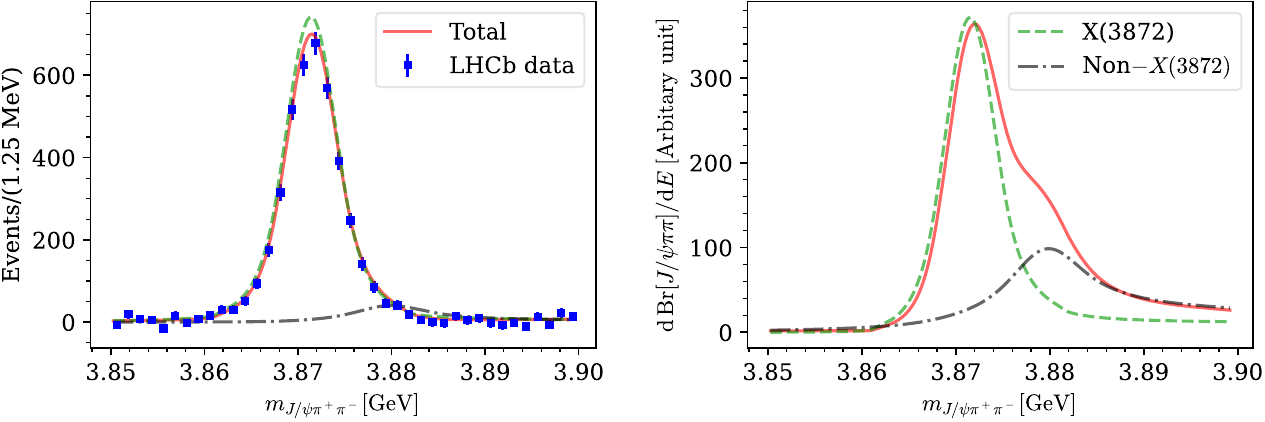}
	\caption{Fit to the LHCb data for $B^+\to K^+J/\psi \pi^+\pi^-$ (left panel) and typical prediction
	for  $B^0\to K^0J/\psi \pi^+\pi^-$ (right panel) controlled by one free parameter.
		\label{Fig:B0Bpldecay}}
\end{figure}

In the charged channels this cusp leads to a very small effect on observables~\cite{Zhang:2024fxy},
however, in the neutral channel, thanks to interference with the $X$-pole, it's effect can be sizeable.
In Ref.~\cite{Ji:2025hjw} it is shown that, since $B^+$ decays with higher probability to $K^+D^0\bar D^{*0}$
than to $K^+D^+D^{*-}$ while this feature is reversed for 
$B^0$ decays, the line shapes for $B\to KJ/\psi \pi\pi$ should be very different for the decays
of these two $B$-mesons as shown in Fig.~\ref{Fig:B0Bpldecay}:
The left panel shows the fit to the $B^+$ decay data, while
 the right panel shows a typical prediction for the line shape in the $B^0$ decay---the exact line shape depends on 
the relative importance for the production of the charged and neutral $D\bar D^*$ pair, which is not known.

The central findings presented in this part are:
\begin{itemize}
\item we demonstrated that, if we insist on isospin violation of natural size for the couplings, the existing high precision 
data can be described in the molecular picture of the $X(3872)$ only, if at the same time there exists
an additional state close to the $D^+D^{*-}$ threshold;
\item this pole is naturally located on a hidden sheet and therefore influences the physical axis and
thus data only exactly at the $D^+D^{*-}$ threshold;
\item this pole should show up visibly in the $X$ line shape extracted from $B^0\to K^0J/\psi \pi^+\pi^-$, although
a quantitative prediction is not possible, since the source strengths are not known.
\item What is listed for the molecular picture is very different to what is found in the diquark--anti-diquark picture: while
also there a pair of states with some mixture of isospin 1 and 0 emerges, the pair is located close together and thus
there is no pole close to the $D^+D^{*-}$ threshold in this scenario~\footnote{For a study where these pictures are
mixed, see Ref.~\cite{Germani:2026yhe}.}.
 
\end{itemize}

\section{Summary}

Recent theoretical and experimental progress put us into the position, that resolving the structure of QCD exotics
got into reach. In particular, hadronic effective field theories allow us, to systematically investigate the imprint of 
the approximate QCD symmetries on hadronic observables and to identify classes of observables that are 
sensitive to the molecular structure of certain exotics. Both, the prediction of spin symmetry partners of the 
vector charmonium like states and the isospin partners of the $X(3872)$ are very non-trivial and emerge naturally
from the molecular picture.  

In the years to come the formalism needs to be extended to also allow for investigations that include the interplay
of regular quark states with the exotics. 
The most systematic approaches in this respect to doubly heavy systems are the Born--Oppenheimer effective field theory (BOEFT)
and the Born--Oppenheimer chiral effective field theory (BO$\chi$EFT).
These EFT s exploit the separation between the slow dynamics of the heavy quarks and the fast dynamics of the light quarks and gluons, see~\cite{Braaten:2014qka,Berwein:2015vca,Brambilla:2017uyf,Oncala:2017hop,Soto:2020xpm,Berwein:2024ztx,Brambilla:2026jqz}
and~\cite{Zhang:2026iyr}, respectively.
The latter, so far applied only to a single system, marks an extension of the formalism presented here that will
be exploited further in the future.


\noindent
{\bf Acknowledgments}

\noindent
I would like to thank my collaborators, in particular X.-K. Dong, T. Ji, V. Baru, L. v. Detten., F.-K. Guo, U.-G. Mei\ss ner and A. Nefediev
for the fruitful collaborations that lead to most of the results presented here.
This work is supported in part by the
 Germany's Excellence Strategy -- EXC 3107 -- Project-ID~533766364. In addition, I thank the CAS President's International Fellowship Initiative (PIFI) under Grant No. 2025PD0087 for partial support.

\bibliography{reference}

\end{document}